\documentclass[prl,aps,twocolumn,showpacs,amsmath,amssymb,floatfix]{revtex4}
\usepackage{bm}
\usepackage{graphicx}
\usepackage{amsmath}
\usepackage{bbold}
\usepackage{dcolumn}
\newcommand{\beq}{\begin{equation}}
\newcommand{\eeq}{\end{equation}}
\newcommand{\beqa}{\begin{eqnarray}}
\newcommand{\eeqa}{\end{eqnarray}}

\begin{document}

\title{\hfill{\small {\bf MKPH-T-04-11}}\\
{\bf Spin Asymmetry and Gerasimov-Drell-Hearn Sum Rule for the Deuteron.}
}
\author{Hartmuth Arenh\"ovel, Alexander Fix and Michael Schwamb}
\affiliation{Institut f\"ur Kernphysik, Johannes Gutenberg-Universit\"at, 
D-55099 Mainz, Germany}

\date{\today}

\begin{abstract}
\noindent
An explicit evaluation of the spin asymmetry of 
the deuteron and the associated GDH sum rule is presented which includes  
photodisintegration, single and double pion and eta production as well. 
Photodisintegration is treated with a realistic retarded potential and
a corresponding meson exchange current. For single pion and eta production 
the elementary operator from MAID is employed whereas for double pion
production an effective Lagrangean approach is used. A large cancellation
between the disintegration and the meson production channels 
yields for the explicit GDH integral a value of 27.31~$\mu$b 
to be compared to the sum rule value 0.65~$\mu$b.
\end{abstract}
 
\pacs{11.55.Hx, 13.60.Le, 24.70.+s, 25.20.Lj}
 
\maketitle
 
\section{Introduction}
\label{sec1}
In recent years the interest in the Gerasimov-Drell-Hearn sum 
rule~\cite{Ger65,DrH66} has been revived and 
has become subject to quite intensive research~\cite{GDH2000,GDH2002,DrT04}.
This sum rule links 
the anomalous magnetic moment $\kappa$ of a particle to 
the integral over the 
energy weighted spin asymmetry of the absorption cross section 
with respect to circularly polarized photons and a polarized target
\begin{eqnarray*}
I^{GDH}=\int_0^\infty \frac{d\omega}{\omega}
\Big(\sigma^P(\omega)-\sigma^A(\omega)\Big)
\,&=&\,\, 4\,\pi^2 \kappa^2\frac{e^2}{M^2}\,S\,,
\end{eqnarray*}
where $S$ denotes the spin of the particle and $M$ its mass.

Obviously, for $\kappa \neq 0$ the particle possesses 
an internal structure. However, the opposite is not in general true. 
A particle having a vanishing or very small $\kappa$ need not be
pointlike or nearly pointlike. In this respect, the deuteron is a
particularly instructive example because it 
has a very small anomalous magnetic
moment $\kappa_d \,=\,-\,0.143$~n.m. and thus a very small sum rule value
$I^{GDH}_d\,=\, 0.65\, \mu\mbox{b}$. 
On the other hand it is well known, 
that the deuteron has quite an extended spatial structure due to its 
small binding energy. The smallness of $\kappa_d$ arises from an almost
complete cancellation of the anomalous magnetic moments of neutron and 
proton because their spins are parallel and predominantly aligned along
the deuteron spin. 

Therefore, it is expected that also for the sum rule integral 
such cancellation of different contributions occurs. 
Indeed, it has been shown in previous 
work~\cite{ArK97,Are01} that a large negative contribution to the sum rule
of $-413~\mu$b arises from the photodisintegration channel, 
which has its origin in a large negative spin asymmetry right above
break-up threshold. It is almost equal in absolute size 
to the sum of neutron and proton GDH values 
$I^{GDH}_p + I^{GDH}_n\,=\, 438\, \mu\mbox{b}$, the latter value 
taken as a rough estimate of the meson production contributions. 

This cancellation of contributions from quite different energy regions
might be surprising on first
sight. But a closer look reveals indeed an intimate connection via
the underlying strong interaction dynamics because the latter is dominated in
both energy regions by pion degrees of freedom, 
namely at low energy via the dominance of the 
one-pion-exchange potential and at high energy via pion production as 
the primary absorptive mechanism on the nucleon.

However, the previous explicit evaluation in~\cite{ArK97} suffered 
from several shortcomings. First of all, only single pion production was
considered. The production operator was very simple and included besides 
Born terms only the $\Delta$-resonance, and final state interaction (FSI) was 
neglected. Thus, the GDH-integral was evaluated up to 550~MeV only 
where convergence had not been reached. Furthermore, for the 
photodisintegration channel the $\Delta$-excitation 
was treated in the impulse approximation (IA) neglecting the important 
dynamical treatment of the N$\Delta$-configuration. 
As a result, a value of $-187$~$\mu$b for the total GDH-integral up to 550 MeV 
was obtained to be compared to the sum rule value 
$I^{GDH}_d\,=\, 0.65\, \mu\mbox{b}$. In subsequent work~\cite{DaA03}, the
treatment of single pion production has been improved by including complete 
FSI in the $\pi$N- and NN-subsystems. However, the simple operator form was 
kept and thus the limited integration range. The major FSI effect was a
reduction of the spin asymmetry of the incoherent $\pi^0$-contribution which 
increased the discrepancy of the explicit calculation to the sum rule value 
even further. 

The scope of the present paper is to present
a considerably improved calculation with respect to photodisintegration
and single pion production, and with additional inclusion of 
two-pion and eta production. 

\section{Spin Asymmetries for the Deuteron}

As elaborated in detail in~\cite{ArK97}, the asymmetry of the 
photodisintegration channel is governed in the near threshold 
region by the dominant isovector M1-transition to the 
$^1$S$_0$-state. Since this state can only be reached for 
antiparallel photon and deuteron spins, a negative 
spin asymmetry arises which is quite huge 
\begin{figure}[htb]
\includegraphics[scale=.6]{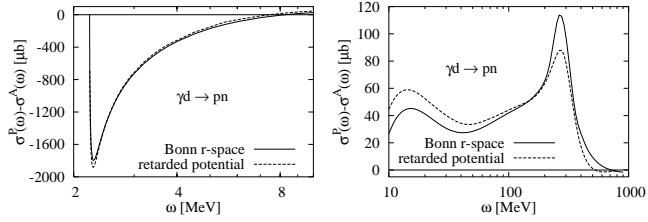}
\caption{Spin asymmetry of deuteron photodisintegration 
using (a) Bonn r-space potential\protect~\cite{MaH87}+MEC+IC+RC and
(b) retarded potential + retarded $\pi$-MEC, 
$\Delta$-degrees in coupled channel, $\pi d$-channel 
+ RC\protect~\cite{ScA01}. Left panel: low energy region; right panel: 
high energy region.}
\label{spin_asy_np}
\end{figure}
\begin{figure}[h]
\includegraphics[scale=.6]{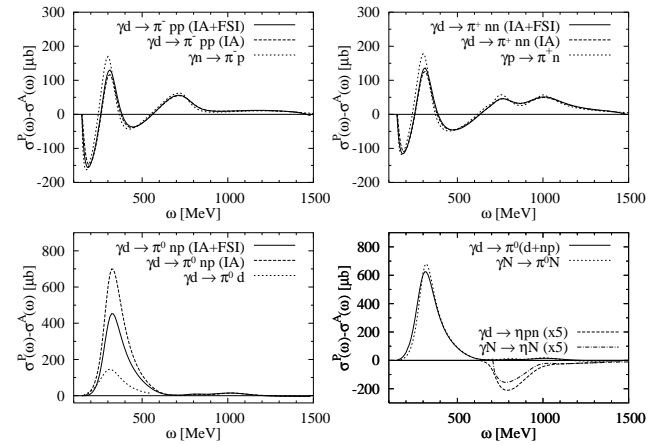}
\caption{Spin asymmetries of single pion and eta production on nucleon 
and deuteron for various charge channels. The results for the deuteron are in 
IA and with inclusion of FSI in final NN- and $\pi$N-subsystems. Upper panels: 
charged single pion production; lower left panel: coherent and incoherent 
$\pi^0$-production on deuteron; lower right panel: total 
$\pi^0$- and $\eta$-production on nucleon and deuteron. For $\pi^0$- and 
$\eta$-production on the nucleon the sum of spin asymmetries of 
neutron and proton is shown. The result for coherent $\pi^0$-production 
is taken from~\cite{ArK97}.}
\label{spin_asy_piNN}
\end{figure}
as is shown in Fig.~\ref{spin_asy_np}. In addition to the earlier 
evaluation~\cite{ArK97}, we show the result of a 
recent considerably improved calculation~\cite{ScA01} which is
based on a retarded potential with retarded 
$\pi$-exchange currents (MEC), incorporation of the N$\Delta$-dynamics (IC)
into a coupled channel approach with retarded interactions, inclusion
of the $\pi d$-channel and relativistic contributions (RC). 
In comparison to~\cite{ScA01} one finds, besides a small increase of the 
asymmetry right at threshold, a significant change at higher energies, 
in particular in the $\Delta$-region. 

For incoherent single pion and eta production on the deuteron an 
improved evaluation of the spin asymmetries is now available in which for 
the elementary production operator the more realistic MAID 
model~\cite{MAID} is used allowing one to extend the calculation to
considerably higher energies. The results for $\pi$-production 
in IA and with inclusion 
of complete rescattering in the final NN- and $\pi$N-subsystems like 
in~\cite{DaA03} are shown in Fig.~\ref{spin_asy_piNN}
together with the corresponding asymmetries for the elementary reactions.
Furthermore, in the lower right panel of Fig.~\ref{spin_asy_piNN}
we show $\eta$-production on nucleon and deuteron, enlarged by a 
factor 5. For the deuteron only incoherent production with NN-FSI 
is shown while coherent production is negligible.
For $\pi$-production one notes besides a positive contribution from 
the $\Delta(1232)$-excitation another one above a photon energy of 
600~MeV from $D_{13}(1520)$ and $F_{15}(1680)$. 
For charged pion production FSI effects are in general quite small, 
except near threshold and in the maximum. The same is true for eta
production. But FSI is quite 
sizeable for incoherent neutral pion production due to the 
non-orthogonality of the final state plane wave in IA to the deuteron
bound state wave function. Comparing the deuteron spin asymmetries 
in Fig.~\ref{spin_asy_piNN} to the corresponding nucleon spin 
asymmetries, one finds a significant 
difference.
\begin{figure}[htb]
\includegraphics[scale=.5]{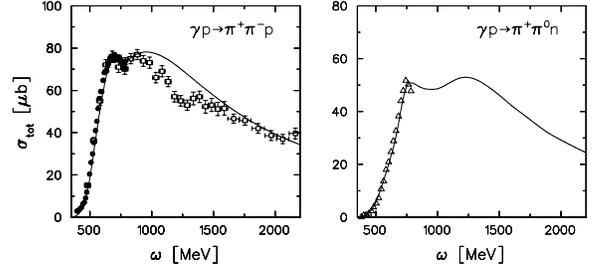}
\caption{Total cross section for the dominant two-pion production
channels on the proton. Experimental data from~\protect\cite{Brag,Lang}.}
\label{2pi_tot}
\end{figure}

\begin{figure}[htb]
\includegraphics[scale=.6]{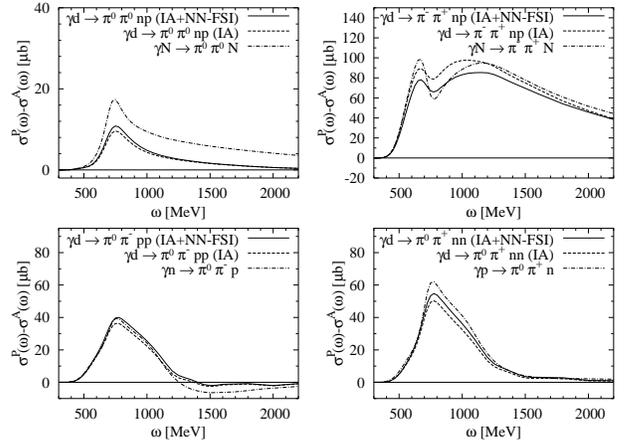}
\caption{Spin asymmetries of double pion production on nucleon 
and deuteron for various charge channels.
For $\pi^0\pi^0$- and $\pi^-\pi^+$-production on the nucleon the sum of 
spin asymmetries of neutron and proton is shown. The deuteron 
results are in IA and with inclusion of FSI in the final NN-subsystem.}
\label{spin_asy_pipi}
\end{figure}

For two-pion production our evaluation is based on a traditional 
effective Lagrangean approach similar to the one in Ref.~\cite{Oset}. 
A detailed description of the 
model including the parameter values and functional form of 
the vertices will be published elsewhere~\cite{FiA04}. The model 
gives a satisfactory description of the available total cross section 
data on the proton as is shown in Fig.~\ref{2pi_tot} for the two 
dominant channels. 

The resulting spin asymmetries for double pion production on the deuteron 
are shown in Fig.~\ref{spin_asy_pipi}, both in IA and with inclusion 
of NN-rescattering (NN-FSI). The latter is small. The largest
contribution comes from the $\pi^-\pi^+$-channel which has not reached 
convergence at 2.2~GeV in contrast to the other three channels.
Furthermore, one notes again significant differences to the elementary 
reaction.

\section{Explicit Evaluation of the GDH Sum Rule for the Deuteron}

\begin{figure}[ht]
\includegraphics[scale=.6]{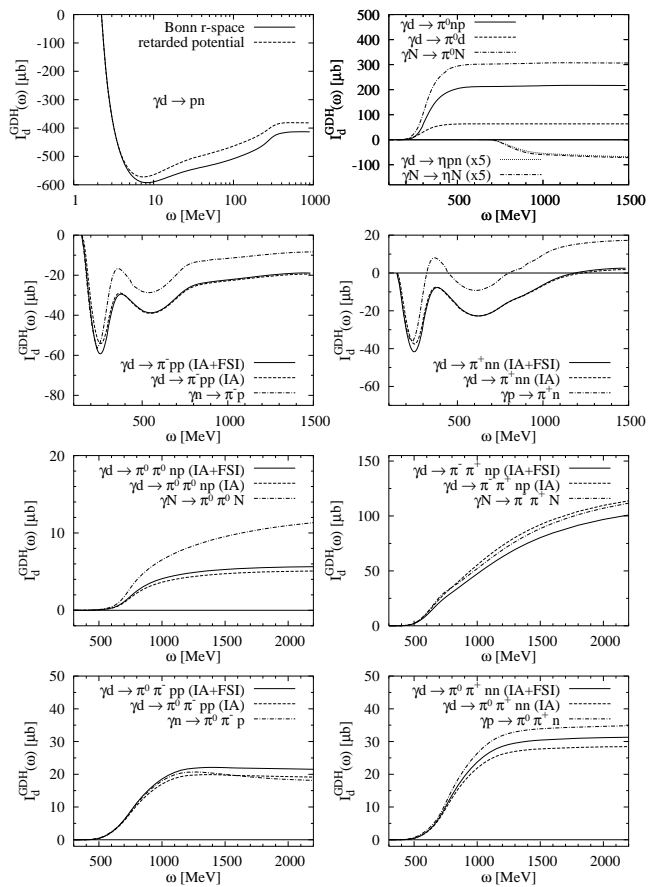}
\caption{Contribution of various channels to the finite GDH integral 
as function of the upper integration limit for deuteron disintegration, 
single and double pion and eta production on nucleon and deuteron. For 
the neutral charge channels $\pi^0$, $\eta$, $\pi^0\pi^0$, and $\pi^-\pi^+$, 
the nucleon integrals are the sum of neutron and proton integrals.}
\label{int_gdh_all}
\end{figure}

Now we will turn to the explicit evaluation of the finite GDH integral
as defined by
\begin{eqnarray*}
I^{GDH}(\omega)=\int_0^\omega \frac{d\omega'}{\omega'}
(\sigma^P(\omega')-\sigma^A(\omega'))\,.
\end{eqnarray*}
The results for photodisintegration, single and double pion 
and eta production are exhibited in Fig.~\ref{int_gdh_all}.
With respect to the photodisintegration channel, one notes a 
significant reduction of the absolute size 
from retardation in potential and $\pi$-MEC at 
lower energies and at higher energies from an improved treatment of 
the $\Delta$-excitation in a coupled channel approach. For single 
and double pion production on the deuteron we show in addition to the IA 
the influence of NN-FSI which gives the most important FSI contribution. 
It is obvious that convergence of the GDH integral is reached for 
$\gamma d\rightarrow np$ already at about 0.8 GeV and for neutral pion
and eta production at 1.5 GeV, however, not completely for charged 
$\pi$-production and also not for double pion production, in particular for 
the channel $\pi^-\pi^+$. 

The comparison of the finite GDH integral for single and double pion 
production on nucleon and deuteron is also shown in
Fig.~\ref{int_gdh_all}. One sees quite clearly 
significant differences between single charged pion production
on nucleon and deuteron because of (i) Fermi motion and 
(ii) to a lesser extent of final state interaction. Also for
double pion production the differences are sizeable and FSI effects
are comparable for some channels. The difference between pion production on 
nucleon and deuteron is also seen in Table~\ref{tab1},
where the separate contributions from the various charge channels of 
pion production to the GDH integral are listed. 

\begin{table}
\caption{Contributions of various charge states of single and double 
$\pi$-production on neutron, proton and deuteron 
to the finite GDH integral (in $\mu$b), integrated up to 1.5 GeV for single 
pion and up to 2.2~GeV for double pion production.}
\begin{ruledtabular}
\begin{tabular}{lccccccc}
 & $\pi^0$ & $\pi^-$ & $\pi^+$ & $\pi^0\pi^0$  & $\pi^+\pi^-$ 
& $\pi^0\pi^-$ & $\pi^0\pi^+$  \\[.5ex]
neutron  & 147.34 & $-8.39$ &      & 6.42 & 57.49 & 18.11 &       \\
proton   & 159.10 &        & 17.28 & 4.91 & 54.18 &       & 34.84 \\
deuteron & 279.87 & $-18.94 $ & $2.51$ & 5.64 & 100.87  & 21.52 & 31.31 \\
\end{tabular}
\end{ruledtabular}
\label{tab1}
\end{table}

\begin{table}[phtb]
\caption{Contributions of various channels to the finite
GDH integral (in $\mu$b), integrated up to 0.8 GeV for 
photodisintegration, 1.5~GeV for single pion and eta production and 2.2~GeV 
for double pion production on nucleon and deuteron.}
\begin{ruledtabular}
\begin{tabular}{lcccccc}
 &  $np$ & $\pi$ & $\pi\pi$ & $\eta$ & $\Sigma$ & sum rule \\[.5ex]
 neutron &           &  138.95 & 82.02 & $-5.77$ & 215.20 & 233.16\\
 proton  &           &  176.38 & 93.93 & $-8.77$ & 261.54 & 204.78\\
 deuteron & $-381.52$ & $263.44$& 159.34 & $-13.95$ & $27.31$ & 0.65\\
\end{tabular}
\end{ruledtabular}
\label{tab2}
\end{table}
The contributions of various channels to the finite GDH-integral for nucleon 
and deuteron are listed in Table~\ref{tab2}. While for the neutron the 
total sum is about 8~\% lower than the sum rule value, it is too large by about
28~\% for the proton. In contrast to our earlier evaluation one now finds 
indeed for the deuteron 
a large cancellation between the contributions of photodisintegration and 
meson production to the GDH sum rule. The sum of all contributions of
27.31~$\mu$b now is positive though somewhat too large. 
However, one should keep in mind that the present theoretical evaluation 
still contains several uncertainties arising from shortcomings of the 
theoretical model and the neglect of contributions at higher energies. 
Probably, the largest uncertainty arises from two-pion production. As a rough 
estimate of this uncertainty we take from Table~\ref{tab2} 
the difference of about 40~$\mu$b between the explicit evaluation for 
neutron plus proton and the corresponding sum rule value. 

\section{Conclusions and outlook}

The GDH sum rule of the deuteron is a very 
interesting observable of its own value
because of a strong anticorrelation between the spin asymmetries of 
low energy photodisintegration and the one of meson production channels
at higher energies. For the photodisintegration channel 
the spin asymmetry is sensitive to relativistic contributions, 
in particular to the spin-orbit current, to meson retardation 
in potential and associated two-body currents, and to a dynamical
treatment of the $\Delta$-excitation. Retardation and N$\Delta$-dynamics 
result in a reduction by about 8~\% of the GDH-contribution 
compared to previous evaluations without such refinements. An experimental 
test of these features would be highly desirable. For this channel, the 
integral is well converged. The uncertainty introduced by the neglect of 
higher isobar configurations beyond the most important N$\Delta(1232)$ one 
is estimated in IA~\cite{ScA95} to be of the order of 5~$\mu$b.

The explicit evaluation of the contributions from single pion and 
eta production to the GDH-integral up to 1.5~GeV and for double pion 
up to 2.2~GeV results in
a positive contribution about equal in absolute size than the one from
photodisintegration confirming thus the large cancellation between 
photodisintegration and meson production as required by the tiny 
sum rule value. However, the resulting total value of the finite integral 
of 27.31~$\mu$b overshoots the sum rule value. Moreover, not all channels of
single and double pion production had reached complete convergence.
In addition, one should keep in mind that the interaction effects 
of meson production are taken into account in an approximate manner 
only. Another uncertainty, though probably quite small, 
arises from the neglect of three-pion, kaon etc.\ 
production. Thus, there is room for improvements of the theoretical 
framework which will allow one to close the gap between the explicit 
evaluation and the sum rule value.

The strong cancellation between the regions at low and high 
energies is a fascinating feature clearly demonstrating the decisive 
role of the pion as a manifestation of chiral symmetry governing 
strong interaction dynamics in these two different energy regions.
The cancellation constitutes also a challenge for any theoretical 
framework since 
it requires a unified consistent treatment of hadron and e.m.\ 
properties for both energy regions.

With respect to meson production channels on nucleon and deuteron,
the corresponding spin asymmetries show a significantly different
behaviour. This means that a direct experimental access to 
the neutron spin asymmetry from a measurement of the spin
asymmetry of the deuteron by subtracting the one of the free proton
is not possible. On the other hand, polarization data of meson
production on the deuteron certainly will provide 
a more detailed test of meson production on the neutron and thus, 
in an indirect manner, on its spin asymmetry. 
However, for this endeavour a reliable theoretical model is needed.
First significant steps in this direction 
have been presented in the present work.

\acknowledgements{This work was supported by the Deutsche 
Forschungsgemeinschaft (SFB 443).}


\begin{thebibliography}{0}
\bibitem{Ger65} S.B. Gerasimov, Yad. Fiz. {\bf 2}, 598 (1965)
                (Sov. J. Nucl. Phys. {\bf 2}, 430 (1966)).
 
\bibitem{DrH66} S.D. Drell and A.C. Hearn, Phys. Rev. Lett.\
                {\bf 16}, 908 (1966).

\bibitem{GDH2000}
Proc. Symposium on the GDH sum rule,
	Mainz 2000, eds. D. Drechsel and L. Tiator (World Scientific,
	Singapore 2001).

\bibitem{GDH2002}
Proc. Symposium on the GDH sum rule,
	Genova 2002, eds. M. Anghinolfi, M. Battaglieri, and R. de Vita 
	(World Scientific, Singapore 2003).

\bibitem{DrT04}
D. Drechsel and L. Tiator, nucl-th/0406059.

\bibitem{ArK97} 
H. Arenh\"ovel, G. Kre\ss, R. Schmidt, and P. Wilhelm,
Phys. Lett. {\bf B407}, 1 (1997).

\bibitem{Are01} 
H. Arenh\"ovel, Ref.~\cite{GDH2000}, p.\ 67.

\bibitem{DaA03}
E.M. Darwish, H. Arenh\"ovel, and M. Schwamb, 
Eur. Phys. J. A {\bf 17}, 513 (2003).

\bibitem{MaH87} R. Machleidt, K. Holinde, and Ch. Elster, 
                Phys. Rep. {\bf 149}, 1 (1987). 

\bibitem{ScA01} 
M. Schwamb and H. Arenh\"ovel, Nucl. Phys. {\bf A690}, 682 (2001).

\bibitem{MAID}
D. Drechsel, O. Hahnstein, S.S. Kamalov, and L. Tiator, Nucl. Phys. {\bf A645},
145 (1999).

\bibitem{Oset}
J.A. Gomez Tejedor and E. Oset, Nucl. Phys. A {\bf 600}, 413 (1996). 

\bibitem{FiA04} 
A. Fix and H. Arenh\"ovel, to be published. 

\bibitem{Brag} 
A. Braghieri {\it et al.}, Phys. Lett. B {\bf 363}, 46 (1995).

\bibitem{Lang}
W. Langg{\"a}rtner {\it et al.}, Phys. Rev. Lett. {\bf 87}, 052001 (2001).

\bibitem{ScA95} 
M. Schwamb, H. Arenh\"ovel, and P. Wilhelm, 
Few-Body Syst. {\bf 19}, 121 (1995).

\end{thebibliography}
\end{document}